\RequirePackage{fix-cm}
\documentclass[smallextended]{svjour3}       
\smartqed  
\usepackage{graphicx}
\usepackage{epstopdf}
\usepackage{pstricks}
\newcommand{\beq}{\begin{equation}}
\newcommand{\eeq}{\end{equation}}
\newcommand{\bea}{\begin{eqnarray}}
\newcommand{\eea}{\end{eqnarray}}
\newcommand{\nn}{\nonumber}

\newcommand{\fig}{Fig.~}

\newcommand{\tr}{{\rm Tr}}
\newcommand{\bx}{{\bf x}}
\newcommand{\by}{{\bf y}}

\def\lsi{\raise0.3ex\hbox{$<$\kern-0.75em\raise-1.1ex\hbox{$\sim$}}}
\def\gsi{\raise0.3ex\hbox{$>$\kern-0.75em\raise-1.1ex\hbox{$\sim$}}}

\begin{document}

\title{Strong coupling methods in QCD thermodynamics
}


\author{Owe Philipsen         
}


\institute{O. Philipsen \at
              ITP, Goethe University Frankfurt,\\
              Max-von-Laue-Str. 1,
              60438 Frankfurt am Main, Germany \\
              \email{philipsen@itp.uni-frankfurt.de}           
}

\date{Received: date / Accepted: date}

\maketitle

\begin{abstract}
For a long time, strong coupling expansions have not  
been applied systematically in lattice QCD thermodynamics,  in view of the succes of numerical Monte Carlo studies. 
The persistent sign problem at finite baryo-chemical potential, however, has motivated investigations using these methods,
either by themselves or 
combined with numerical evaluations, 
as a route to finite density physics. This article reviews the strategies, by which a
number of qualitative insights have been attained,
notably the emergence of the hadron resonance gas or the 
identification of the onset transition to baryon matter in specific regions of the QCD parameter space.
For the simpler case of Yang-Mills theory, the 
deconfinement transition can be determined
quantitatively even in the scaling region, showing possible prospects for continuum physics.

\end{abstract}

\section{Introduction}
\label{intro}
Quantum Chromodynamics (QCD) at finite temperatures and baryon densities is the theoretical foundation for an understanding
of observed phenomena in a wide range of fields, like the thermal history of the early universe, heavy-ion collision experiments 
or the composition and properties of neutron stars. Since the interactions of QCD are strong on scales up to several GeV, ordinary 
weak coupling perturbation
theory fails for temperatures and baryo-chemical potentials $T, \mu_B\lsi 2$ GeV. 
The non-perturbative treatment by numerical Monte Carlo simulations
of lattice QCD is to date still restricted to vanishing or small baryo-chemical potentials, because of
a severe sign problem at finite baryon density (for an introduction, see \cite{Philipsen:2010gj}). 

The continued absence of a genuine algorithmic solution to this problem has motivated renewed interest in strong coupling 
and hopping expansion methods over the last decade. 
Contrary to weak coupling expansions, which typically result in asymptotic series, 
strong coupling expansions in the Euclidean framework are known to yield convergent series with well-defined radii of convergence. 
In the early days of lattice gauge theory they were used to get analytical results for some physical quantities of interest, such as glueball 
masses \cite{Munster:1981es,Seo:1982jh} 
or the energy density of lattice Yang-Mills theories \cite{Balian:1974xw,Drouffe:1983fv}.
These calculations were restricted to zero temperature, with the exception of
some mean field analyses of phase transitions in the strong coupling limit  \cite{Polonyi:1982wz,Green:1983bq,Green:1983sd,Gocksch:1984xc}, 
and a series for the temperature-dependent string tension \cite{Green:1982hu}.
More recent attempts to explore strong coupling methods as a systematic calculational tool for thermodynamics started with 
an investigation of Yang-Mills theory \cite{Langelage:2008dj}.
While Monte Carlo methods appear superior where they work, analytic results provide additional understanding
for the interpretation of data. Moreover a two-stage approach, with an analytic derivation of effective lattice theories and their
subsequent analytic or numerical solution, appears to be reasonably promising for finite baryon density. It is currently the only way to
investigate the onset transition to baryon matter in lattice QCD directly, albeit not yet for physical parameter values.
The purpose of this article is to review these developments with a focus on the concepts and the results, whereas technical
details are left to the references.

\section{A proof of concept: the $SU(3)$ spin model}
\label{sec:su3spin}

Before approaching the difficulty and complexity of QCD, it is useful to illustrate the basic strategies at the example of
a simpler theory that is fully understood and has served several times as a testing ground for methods to 
deal with the sign problem. We consider the $SU(3)$ spin model with the action
\bea\label{eq:su3spinaction}
S=-\sum_x \Big(\sum_{k=1}^3\tau \left[L(x)L^*(x+\hat{k})+L^*(x)L(x+\hat{k})\right]
+\eta L(x)+\bar{\eta}L^*(x)\Big).
\eea
The fields $L(x)=\tr W(x)$ are complex scalars representing the trace of $SU(3)$-matrices $W(x)$. Writing
$\eta=\kappa\exp(\mu), \bar{\eta}=\eta(-\mu)$, the model closely
resembles an effective action of lattice QCD with static quarks in the strong coupling region, to be discussed in Sec.~\ref{sec:heavy}.   
 For $\mu=0$, it features spontaneous breaking of its $Z(3)$ center symmetry 
along a line $\tau_c(\kappa)$ of first-order transitions with an endpoint, as well as a sign problem at 
$\mu\neq 0$.
The model can be reformulated free of a sign problem in terms of a flux 
representation \cite{Gattringer:2011gq}, 
allowing for simulations \cite{Mercado:2012ue} by means of a worm algorithm. 
Likewise, a complex Langevin  
algorithm has been successfully applied \cite{Karsch:1985cb,Bilic:1987fn,Aarts:2011zn}. 

\begin{figure}
  \includegraphics[width=0.5\textwidth]{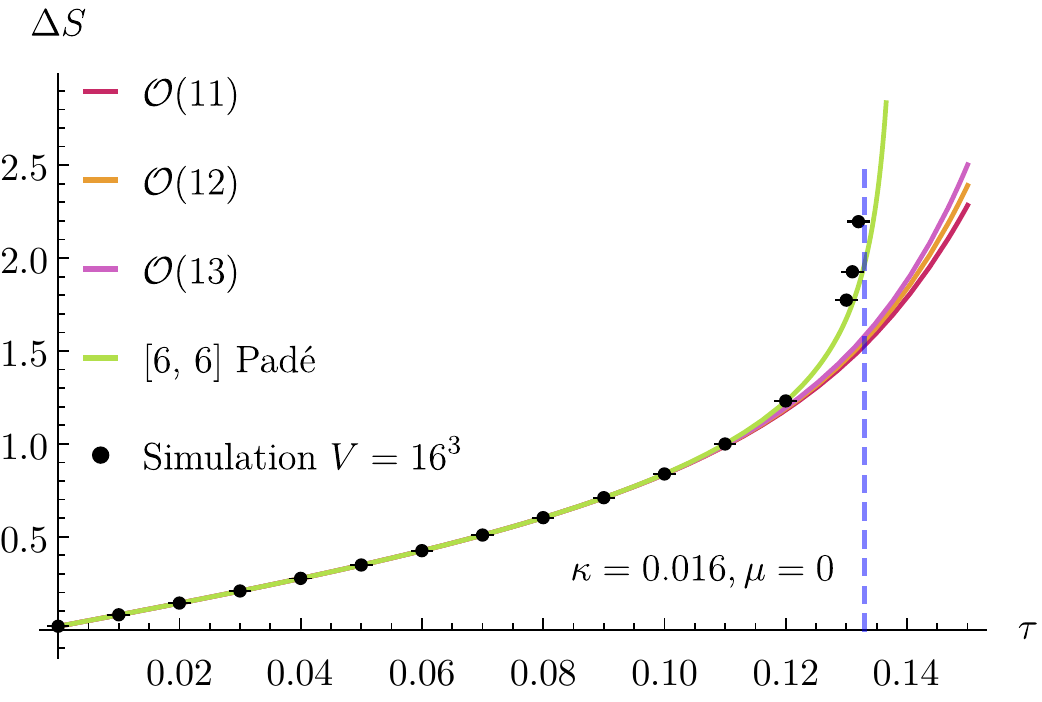} \hspace*{0.3cm}
  \includegraphics[width=0.5\textwidth]{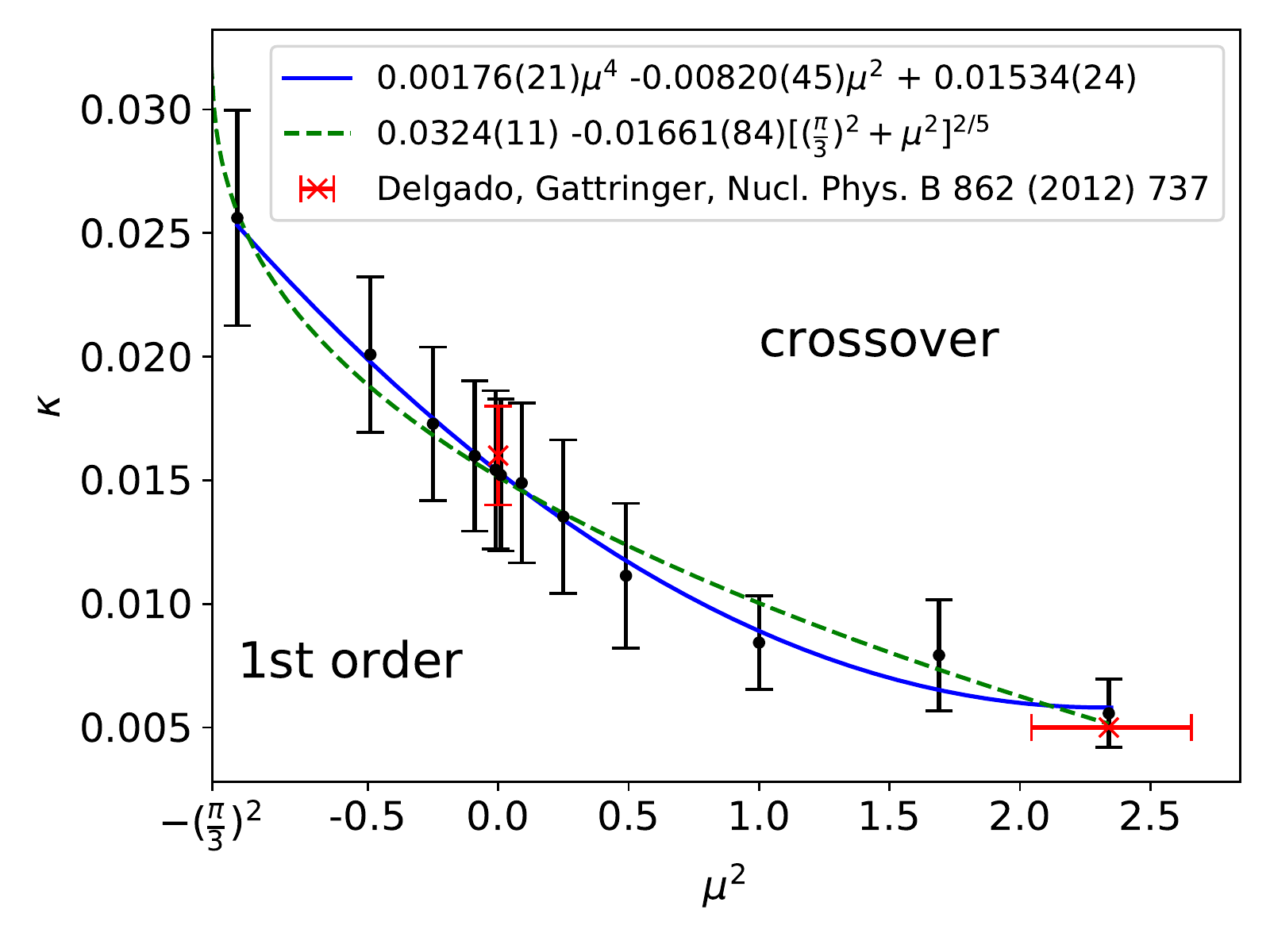}
\caption{Equation of state from the linked cluster expansion compared to Monte Carlo. The vertical line marks the center breaking phase transition.
Right: Order of the phase transitions, at an implicit $\tau_c(\kappa,\mu)$, in the $SU(3)$ spin model. First-order and crossover regions are separated by a critical line with 3d $Z(2)$ universality.   
From \cite{Kim:2020atu}.}
\label{fig:su3_spin}       
\end{figure}

Here, we are interested in a solution by series expansion methods, which are more familiar in condensed matter physics.
Specifically, consider a linked cluster expansion of the free energy density (for an introduction, see \cite{Wortis:1980zc}), which has been 
evaluated through 14 orders in $\tau$, and for each order up to convergence in $\kappa$ \cite{Kim:2020atu},
\beq
  f = - \frac{\ln(Z)}{V}=\sum_{n=0}^{14} a_n(\kappa,\mu) \tau^n\;, \quad \Delta S  = - \frac{\partial f}{\partial \tau} - \frac{\partial f}{\partial \eta}\;.
 \eeq
 The equation of state can be parametrised by the combined derivatives $\Delta S$, corresponding to the interaction
 measure in QCD.
It can be straightforwardly determined in Monte Carlo simulations, and all thermodynamic
functions follow by integration or differentiation. Results corresponding to the three highest 
orders are shown in \fig\ref{fig:su3_spin} (left). Excellent convergence and agreement with Monte Carlo results is observed
until the phase transition, representing the radius of convergence, is approached. A finite polynomial is unable to describe
a non-analytic phase transition, but is sensitive to it
by loss of convergence.
 
A marked improvement near the transition
can be obtained by infinite-order resummations through Pad\'e approximants, defined as rational functions,
\beq
[L,M](x)\equiv\frac{a_0+a_1x+\ldots+a_L x^L}{1+b_1x+\ldots+b_M x^M}\;.
\eeq
The coefficients $a_i, b_i$ are uniquely determined for $L+M \leq N$, if $N$ represents the highest available order 
of the expansion. In this way the
$[L,M]$ approximant reproduces the known series up to and including $\mathcal{O}(x^{L+M})$. Pad\'e approximants are able to show
singular behaviour and can model scaling properties near second-order phase transitions. The $[6,6]$ approximant to $\Delta S$ is 
shown also in \fig\ref{fig:su3_spin} (left). It accurately reproduces the simulated equation of state all the way to the phase transition, which it 
indicates by a singularity. 

At $\mu=0$ the phase transition is known to be first-order for small $\kappa$, ending in a critical point at some $\kappa_c$, which is
of the same type as standard liquid-gas transitions.
Therefore the susceptibility and the specific heat,
 \beq
  \chi =
  - \frac{\partial^2 f}{\partial \eta^2}
  - \frac{\partial^2 f}{\partial \bar{\eta}^2}
  - 2 \frac{\partial^2 f}{\partial \eta \partial \bar{\eta}}\;,\quad
  C=-\frac{\partial^2 f}{\partial\tau^2}\;,
  \label{eq:specheat}
\eeq
when approaching from any direction not parallel to the scaling axes,
must both diverge with a common critical exponent, $\chi\sim C\sim |\tau- \tau_c|^{-\epsilon}$, $\epsilon=\gamma/\beta\delta$. 
This is reflected in a simple pole of the Dlog-Pad\'e's, e.g.,
 \beq
 D_\chi(\tau)\equiv \frac{d}{d\tau}\log \chi\sim -\frac{\epsilon}{(\tau-\tau_c)}\;.
\eeq
A difficulty is that Pad\'e's also show poles for first-order transitions, where they indicate the end of the metastability range, 
as well as spurious
poles, which renders the analysis quite intricate. 
In order to detect a true critical point one has to require convergence of the poles predicted by different Pad\'e's,
as well as convergence of he poles in both observables, $D_\chi,D_C$. In this way, a critical point can be located within some
scatter specifying a systematic error. Once the series are long enough to achieve this, an application to finite chemical potential, real
or imaginary, poses no additional problem. Hence, the phase structure and the order of the phase tansition can be fully determined,
as shown in \fig\ref{fig:su3_spin} (right), in quantitative agreement with numerical results. This example provides further 
motivation to explore 
such methods also in QCD, where no algorithmic solutions to the sign problem exist yet.

\section{Strong coupling and hopping expansions in QCD at finite T}

\subsection{Yang-Mills theory}
\label{sec:ym}

Consider the Yang-Mills partition function with the standard Wilson action
\beq
Z=\int \,DU\,\exp -S_{YM}[U],\quad S_\mathrm{YM}=\sum_p\frac{\beta}{2N_c}\left( \mathrm{Tr}\, U_p+\mathrm{Tr}\,U_p^{\dagger}\right),
\eeq
with plaquette variables $U_p$ and the lattice coupling $\beta=\frac{2N_c}{g^2}$.
From the lattice action it is obvious that the exponential can be expanded in powers of $\beta$, where the expansion
point $\beta=0$ corresponds to the strong (infinite) coupling limit. 
However, in the continuum limit $\beta\rightarrow \infty$ and convergence is expected to be slow. 
A powerful resummation
including all orders in $\beta$ is immediately achieved by expanding instead in the group characters $\chi_r(U_p)=\tr D_r(U_p)$, which are
traces of the representation matrices $D_r$ of the plaquette variables. Using the formalism of moments and 
cumulants \cite{Drouffe:1983fv,Montvay:1994cy}, it can be shown that this expansion exponentiates, such that a series 
for the (formal) free energy density  
 $\tilde{f}\equiv-\frac{1}{\Omega}\ln Z$ is obtained in terms of clusters $C$ of graphs $\Phi$,
\bea
\tilde{f}&=&-6\ln\,c_0(\beta)-\frac{1}{\Omega}
\sum_{C=(X_i^{n_i})}\,a(C)\prod_i\Phi(X_i)^{n_i}\;,\nn\\
\Phi(X_i)&=& \int DU\prod_{p\in X_i} d_{r} a_{r}(\beta) \chi_{r}(U_p)\;.
\label{free}
\eea
Here $\Omega=V\cdot N_\tau$ is the lattice volume, $d_r$ and $a_r(\beta)$ are the
dimension and expansion coefficient of representation $r$, and $c_0$ is 
the expansion coefficient of the trivial representation. 
The combinatorial factor $a(C)$  
equals $1$ for clusters $C$ consisting of only one 
so-called polymer $X_i$, which represent closed surfaces of plaquettes. 
The coefficients of higher representations can be expressed in terms of $u(\beta)\equiv a_f(\beta)=\beta/18+\beta^2/216+\ldots$,  
the coefficient of the fundamental representation.  It is a known function over the entire range of lattice gauge couplings with $u(\beta)\in [0,1)$,
and constitutes the expansion parameter for the character expansion. 

For thermodynamics the physical temperature $T=1/(aN_\tau)$ is
realised by compactifying the temporal extension of the lattice. 
The physical free energy is then obtained by subtracting  
the divergent vacuum contribution,
\begin{equation}
f(N_\tau,u)=\tilde{f}(N_\tau,u)-\tilde{f}(\infty,u),
\end{equation}
and the pressure is $P=-f$.
Group integrals are evaluated using the formulae
\begin{equation}
\int dU \chi_r(UV)\chi_r(WU^{-1})=\chi_r(VW),\quad \int dU \chi_r(U)=\delta_{r,0},
\end{equation}
the latter enforces contributing graphs $X_i$ to be objects with a closed surface. 

\begin{figure*}
  \includegraphics[width=0.9\textwidth]{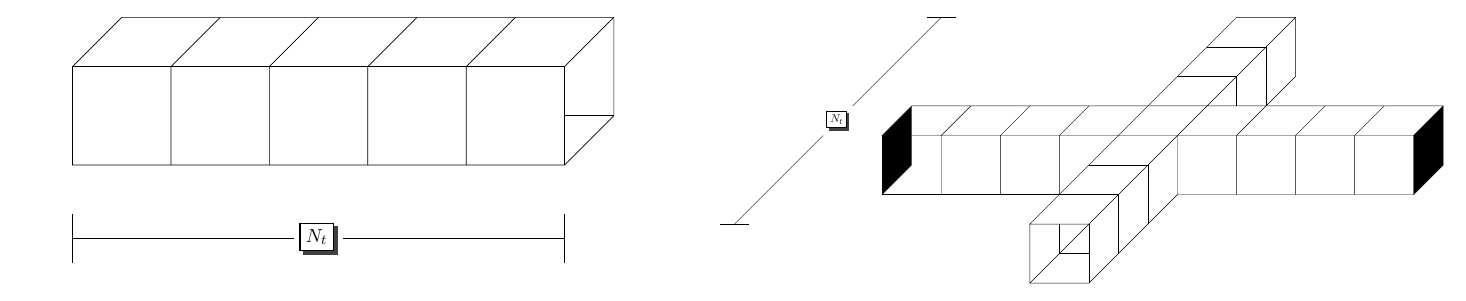}
\caption{Leading order graphs for the free energy density (left) and screening mass (right).}
\label{fig:graphs}       
\end{figure*}

\begin{figure}
  \includegraphics[width=0.5\textwidth]{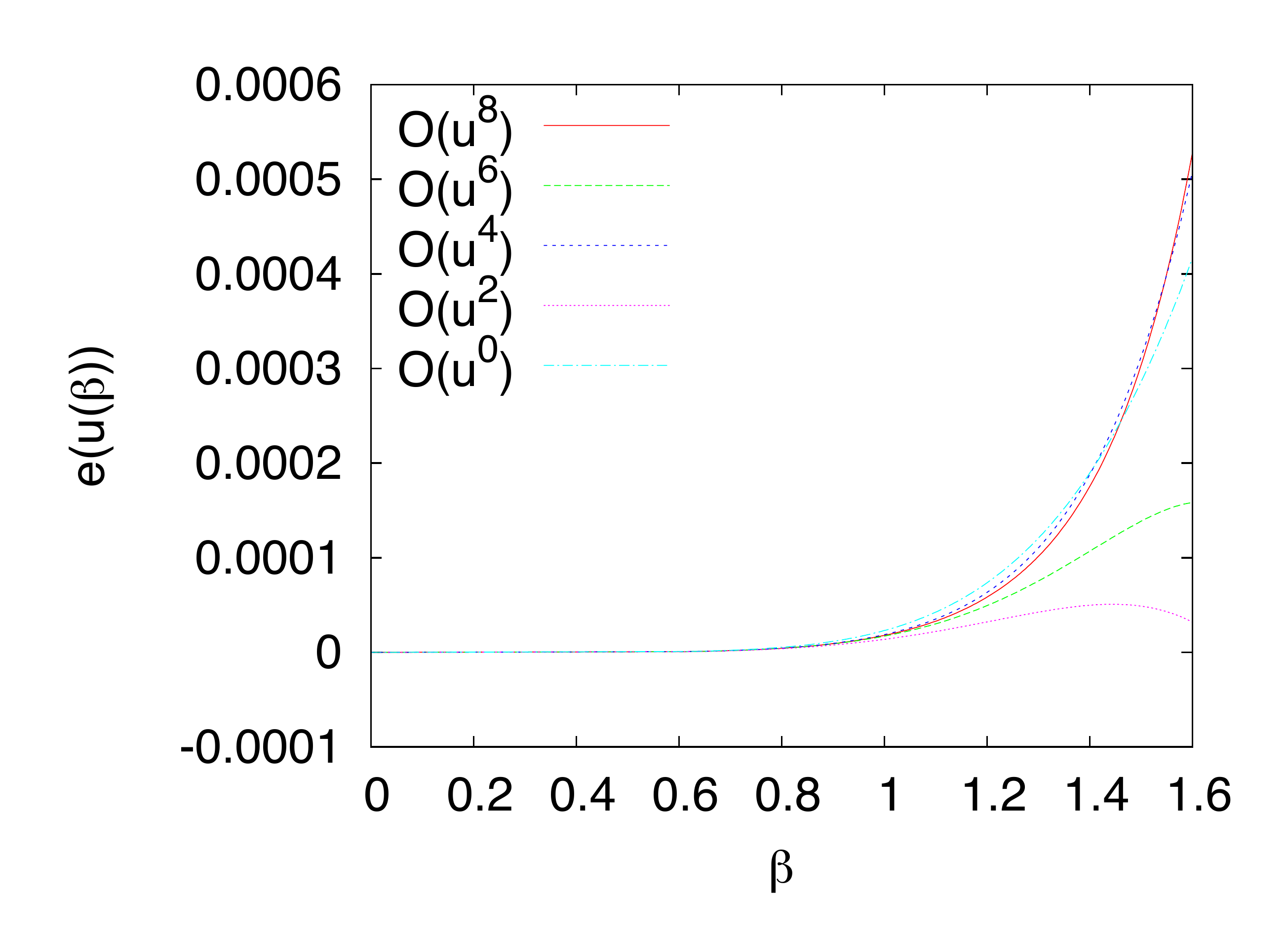} \hspace*{0.3cm}
  \includegraphics[width=0.52\textwidth]{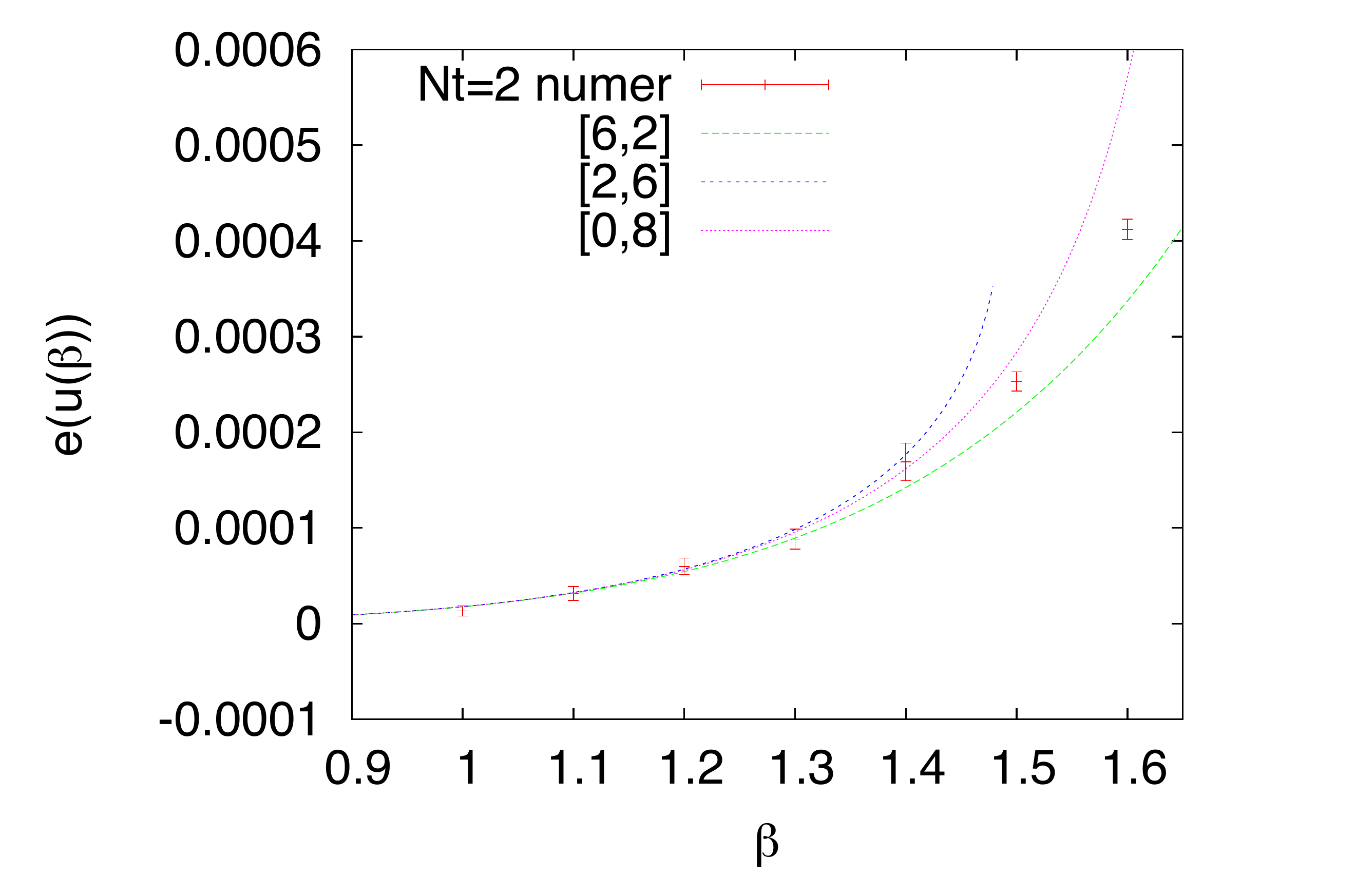}
\caption{Left: Energy density to different orders in the character expansion. Right: Resummation by Pad\'e approximants
and comparison with Monte Carlo.
From \cite{Langelage:2008dj}.}
\label{fig:su2}       
\end{figure}

The leading order graph is a tube of length $N_\tau$, Fig.~\ref{fig:graphs}. 
Summing over all such graphs on the lattice, their contribution to the pressure is 
\beq
P^{LO}(N_\tau,u)=\frac{6}{N_\tau}u^{4N_\tau}\;.
\eeq
Thus, the strong coupling limit at $\beta=0$ (and thus $T=0$) has zero free enery density or pressure,
as expected. Moreover, we recognize a qualitative feature known from full lattice simulations, where it is difficult to extract, 
and hadron resonance gas
descriptions, namely the exponentially slow rise of the pressure with $T$. 
In order to get more quantitative,
corrections to the free energy density through
order $u^8$ with respect to the leading order term for various $N_\tau$ have been
calculated for SU(2) \cite{Langelage:2008dj} and for SU(3) \cite{Langelage:2009jb,Langelage:2010yn}. A direct comparison with Monte Carlo
simulations is  easiest for the energy density, 
\beq
e(\beta)=\frac{1}{6}\frac{d}{d\beta}f(\beta)
=\langle\,{\rm Tr}\,U_p\,\rangle_{N_\tau}-\langle\,{\rm Tr}\,U_p\,\rangle_{N_\tau=\infty}, 
\eeq
which is shown in Fig.~\ref{fig:su2} for the example of $SU(2)$. On the left, 
the series is shown order by order in the  character coefficient $u(\beta)$, and evaluated as a function of $\beta$.
The series loses convergence as it approaches the deconfinement transition,  whose critical coupling $\beta_c$
bounds the radius of convergence in this case. On the right, a resummation of the series is achieved by Pad\'e approximants.
This clearly improves the convergence and comparsion between
different approximants provides an estimate for the systematic error associated with the truncated series. Good agreement
with the Monte Carlo data is observed in the controlled region, which is however restricted to relatively small $\beta$-values.

At the finite temperature phase transition the free energy $f(N_\tau,u(\beta_c))$ is continuous, 
with a discontinuous first or second derivative, depending on the order of the transition.
For $SU(2)$, we have a second-order transition, for which the `heat capacity' 
\beq
C(N_\tau,u)=u^2\frac{d^2}{du^2}f(N_\tau,u)
\eeq
diverges as $C(u)\sim (u_c-u)^\alpha$, with a critical exponent
characteristic of the universality class. Its logarithmic derivative 
\beq
D_C(N_\tau,u)\equiv\frac{d}{du}\ln C(N_\tau,u)\sim -\frac{\alpha}{u_c-u} 
\eeq
has a simple pole with residue $\alpha$ and is therefore well-suited for an analysis
by Pad\'e approximants. The averaged predictions of the approximants in Fig.~\ref{fig:su2} are $\beta_c=1.65(35)$ and
$\alpha=0.052(19)$ \cite{Langelage:2008dj}, compared to the Monte Carlo value $\beta_c=1.880(3)$ \cite{Fingberg:1992ju} 
and the 3d Ising exponent $\alpha=0.12$.

Another interesting qualitative result can be obtained for thermal screening masses, which are
obtained from the exponential decay of spatial correlation functions.
The correlator of plaquettes is given by
\begin{equation}
C(z)=\langle\mathrm{Tr}\,U_{p_1}(0)\,\,\mathrm{Tr}\,U_{p_2}(z)
\rangle=N^2\frac{\partial^2}{\partial\beta_1\partial\beta_2}\ln\,
Z(\beta,\beta_1\beta_2)\bigg\vert_{\beta_{1,2}=\beta}.
\end{equation}
At zero temperature (infinite $N_\tau$) the exponential decay is the same as for correlations in the time direction, and thus
determined by the glueball masses. At finite temperature (finite $N_\tau$)
the LO
graph for the difference to the vacuum mass is shown in 
Fig.~\ref{fig:graphs} (right) and gives
\beq
\Delta m^{LO}(N_\tau,u)=m(N_\tau)-m(\infty)
=-\frac{2}{3}N_\tau u^{4N_\tau-6}\;.
\eeq
Thus finite temperature lowers the screening 
mass in the confined phase, but the effect   
is suppressed by its high order in the expansion parameter. This explains
the insensitiviy of screening masses to $T$ observed
in simulations of the confined phase, where they stay very close to the vacuum hadron masses up
to the phase transition \cite{Datta:2002je}. Such an understanding is not attainable from simulations alone.

\subsection{QCD with heavy quarks}

Including fermions for a treatment of full QCD requires an additional expansion. 
The Wilson fermion determinant can be formulated as a power series in the hopping
parameter $\kappa_f=(2am_f+8)^{-1}$ for every flavour $f$.
Defining the usual hopping matrix $M$ in Dirac, colour and coordinate space,
Grassmann integration over fermion fields gives to leading 
order in $\kappa_f$ per flavour
\begin{eqnarray}
-S_q^f&=&-\tr\,\ln\,(1-\kappa_fM)
=\sum_{l}\,\frac{\kappa_f^l}{l}\,\mathrm{tr}\,M^l(\mu)\nn\\
&=& -(2\kappa_f)^{N_\tau}\sum_{\bf x} \left( e^{\mu N_\tau}L({\bf x})
+e^{-\mu N_\tau}L^\dag({\bf x})\right)
+\ldots 
\label{eq:hop} \\
\mbox{with}\quad M_{ab,\alpha\beta,yx}&=&\sum_{\mu}\delta_{y,x+\hat{\mu}}(1+\gamma_{\mu})_{\alpha\beta}U_{ab,x\mu} \;.
\label{fact}
\end{eqnarray}
The Kronecker deltas in $M$ ensure that only closed loops contribute, and
$l$ counts their length.
Again, for finite temperature effects we only need the 
difference between finite and infinite $N_\tau$, and 
hence are interested in the loops winding through the temporal boundary.
One can formalise this by considering a hopping expansion in the spatial directions only,
while the hops in the Euclidean time direction are taken into account fully. 
Thus the leading order heavy 
fermionic contributions to the effective action can be written as
in Eq.~(\ref{eq:hop}), which only holds for finite $T$, and
the dots represent loops that wind more than once.

Now one can calculate again the free energy density and pressure, 
this time performing character expansions in both $U$ 
and $L$. 
Expanding all terms up to ${\cal{O}}(\kappa^{3N_\tau})$ and 
doing the group integrals one finds for two flavours $u,d$ \cite{Langelage:2010yn}
\begin{eqnarray}
P&=&\frac{1}{N_\tau a^4}\bigg\lbrace4(2\kappa_u)^{2N_\tau}+8(2\kappa_u2\kappa_d)^{N_\tau}+4(2\kappa_d)^{2N_\tau}\bigg\rbrace\nonumber\\
&+&\frac{1}{N_\tau a^4}\bigg\lbrace4(2\kappa_u)^{3N_\tau}+6\big[(2\kappa_u)^22\kappa_d\big]^{N_\tau}\nonumber\\
&+&6\big[2\kappa_u(2\kappa_d)^2\big]^{N_\tau}+4(2\kappa_d)^{3N_\tau}\bigg\rbrace\Big(\mathrm{e}^{3a\mu N_\tau}
+\mathrm{e}^{-3a\mu N_\tau}\Big)\;.
\end{eqnarray}
Next we identify the hadron masses to leading order in the hopping
expansion,
\bea
\mbox{Mesons:}\qquad\quad am_{f\bar{f^\prime}}&=&-\ln2\kappa_f-\ln2\kappa_{f^\prime}\\
\mbox{Baryons:}\qquad am_{ff^\prime f^{\prime\prime}}&=&-\ln2\kappa_f-\ln2\kappa_{f^\prime}-\ln2\kappa_{f^{\prime\prime}}.
\eea 
The pressure can then be rewritten as
\begin{eqnarray}
P&=&\frac{1}{N_\tau a^4}\left\lbrace\sum_{0^-}\mathrm{e}^{-m\left(0^-\right)N_\tau}+3\sum_{1^-}\mathrm{e}^{-m\left(1^-\right)N_\tau}\right\rbrace\nonumber\\
&+&\frac{1}{N_\tau a^4}\Bigg\lbrace4\sum_{\frac{1}{2}^+}\mathrm{e}^{-m\left(\frac{1}{2}^+\right)N_\tau}+8\sum_{\frac{3}{2}^+}
\mathrm{e}^{-m\left(\frac{3}{2}^+\right)N_\tau}
\Bigg\rbrace\cosh\big(\mu_BN_\tau \big),
\label{eq_hrg}
\end{eqnarray}
which is the expression corresponding  to an ideal gas of hadrons.
We have thus derived from first principles that QCD
reduces to a hadron resonance gas in the strong coupling and heavy mass regime.
Note that this also holds for Yang-Mills
theory, which can be represented as a glueball gas \cite{Langelage:2008dj,Langelage:2010yn}.
It is then plausible that this feature is generic and independent of quark mass, which 
explains the success of hadron resonance gas descriptions of the lattice QCD equation of state 
in the confined phase, where the physical gauge coupling is fairly strong \cite{Borsanyi:2010cj,Bazavov:2014pvz}.

\subsection{Effective lattice theories from strong coupling methods}

In the previous section we have seen that strong coupling methods lead to interesting qualitative insights in QCD thermodynamics,
but they are also limited quantitatively if we are interested in continuum physics. This can be improved significantly
by a procedure in two stages: 1) construct an effective lattice theory by 
integrating out part of the degrees of freedom, and 2) solve the effective theory by series expansion or otherwise. 
Two types of effective degrees of freedom arise naturally, depending on the integration order,
\beq
Z=\int DUD\bar{\psi}D\psi\;e^{-S_\mathrm{QCD}[U,\bar{\psi},\psi]}=\int DU_0\;e^{-S_\mathrm{eff}[U_0]} 
=\int D\bar{\psi}D\psi\;\;e^{-S_\mathrm{eff}[\bar{\psi},\psi]}\;.
\label{eq:seff}
\eeq
In the first case, fermions as well as all spatial link variables are integrated over, leaving a theory of temporal links only, which
on a periodic lattice can be expressed by Polyakov loops. In the second case, all gauge links are integrated, leaving a fermionic
effective theory in terms of mesons and baryons, because of gauge invariance. Both representations are
perfectly equivalent to QCD. Truncations involved in doing the integrations reduce this equivalence 
to specific parameter regions, where the corresponding approximations hold.

\section{Effective lattice theory for QCD with heavy quarks}
\label{sec:heavy}

The representation of QCD in terms of Polyakov loops was first developed to characterise the 
deconfinement phase transition in pure gauge theories \cite{Polonyi:1982wz,Svetitsky:1982gs}, and has later been considered
perturbatively and non-perturbatively 
in the continuum \cite{Herbst:2013ufa,Vuorinen:2006nz,Fischer:2013eca,Fischer:2014vxa,Lo:2014vba} and on the lattice
\cite{Wozar:2007tz,Greensite:2013yd,Greensite:2013bya,Greensite:2014isa,Bergner:2015rza}.
Here our interest is in the derivation of the effective theories by strong coupling methods. In this case their
form is uniquely determined and can be systematically worked out and improved. Starting point is Wilson's lattice
formulation, as before. However, instead of the free energy we now compute the effective action defined in the first of Eq.~(\ref{eq:seff}),
\beq
-S_\mathrm{eff}[U_0]\equiv \ln \int DU_i \exp(-S_\mathrm{YM}[U_0,U_i]-S_q^f[U_0,U_i])\;. 
\eeq
This is done again by combining a character expansion of the exponentiated gauge action with a 
hopping expansion of the exponentiated quark action. The gauge integrals can then be done term by term
for the truncated expansion to leave an effective action depending on temporal links only. These combine to temporal Wilson lines $W(\bx)$, which
close through the periodic boundary and
implicitly contain the dependence on the Euclidean time extent $N_\tau$.
As a result, the effective partition function is three-dimensional, resembling
a continuous spin model \cite{Langelage:2010yr,Fromm:2011qi},
\bea
\label{zpt}
\label{eq:zeff}
Z&=&\int DW\prod_{<\bx, \by>}\left[1+\lambda(L_{\bx}L_{\by}^*+L_{\bx}^*L_{\by})\right]\\
&&\hspace*{-0.75cm}
\times\prod_{\bx}[1+h_1L_{\bx}+h_1^2L_{\bx}^*+h_1^3]^{2N_f}[1+\bar{h}_1L^*_{\bx}+\bar{h}_1^2L_{\bx}+\bar{h}_1^3]^{2N_f}
\nonumber\\
\hspace*{-0.5cm}
&&\hspace*{-0.72cm}\times \prod_{<\bx, \by>}\Big(1-h_{2}{\rm Tr} \frac{h_1W_{\bx}}{1+h_1W_{\bx}}{\rm Tr} \frac{h_1W_{\by}}{1+h_1W_{\by}}\Big)
\Big(1-h_{2}{\rm Tr} \frac{\bar{h}_1W^\dag_{\bx}}{1+\bar{h}_1W^\dag_{\bx}}{\rm Tr} \frac{\bar{h}_1W^\dag_{\by}}{1+\bar{h}_1W^\dag_{\by}}\Big)
\nn\\
&&\hspace*{-0.72cm}\times \quad \ldots \; \quad .
\eea
Here the first line corresponds to pure gauge theory, the second line to the static quark determinant and the third line to
the leading contributions of the kinetic quark determinant. 
The couplings of the effective theory are functions of the original lattice QCD parameters (for higher order expressions see \cite{Fromm:2011qi}),
\bea
\lambda&=& u^{N_\tau}\exp[N_\tau(4u^4+\ldots)]\;,\nn\\
h_1&=&(2\kappa e^{a\mu})^{N_\tau}\left(1+\ldots \right) =e^{\frac{\mu-m}{T}}(1+\ldots), \quad 
\bar{h}_1=h_1(-\mu)\;,\nn\\
h_2&=&\kappa^2 N_\tau/N_c(1+\ldots)\;.
\label{eq:coupl}
\eea
and $am=-\ln(2\kappa)=am_B/3$ denotes the leading-order constituent quark mass in a baryon (not to be confused with the 
current quark mass $m_q$ in the QCD action). The higher the order to which the effective action is considered, the more
couplings appear which are increasingly non-local, connecting loops $W_{\bf x}$ to all powers and over all lattice separations.
Note that the effective couplings
are (resummed) power series in the original expansion parameters $u$ and $\kappa$, and hence can in turn be treated
as small expansion parameters themselves. Any truncated effective theory 
can thus be treated by series expansion methods as discussed in 
the simple example of Sec.\ref{sec:su3spin}. On the other hand, the effective theory has a much reduced sign problem,
and can be simulated by even a choice of different algorithms. One current line of work is to design flux representations
for this type of actions that can cure the sign problem and be simulated efficiently \cite{Borisenko:2020gjn,Borisenko:2020cjx}.

\subsection{Deconfinement transition in Yang-Mills theory}

\begin{figure}[t]
\centering
\includegraphics[width=0.48\textwidth]{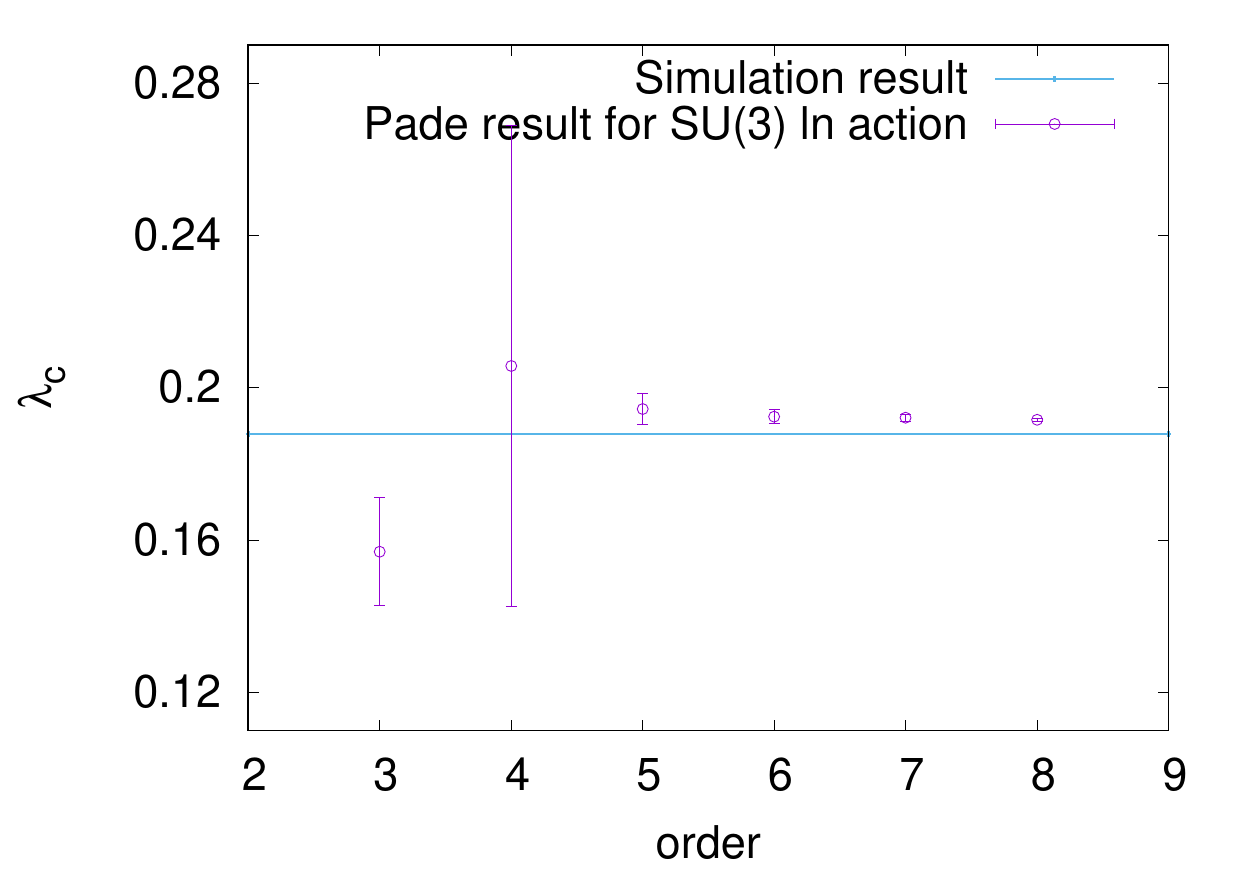}\hspace*{0.5cm}
\includegraphics[width=0.48\textwidth]{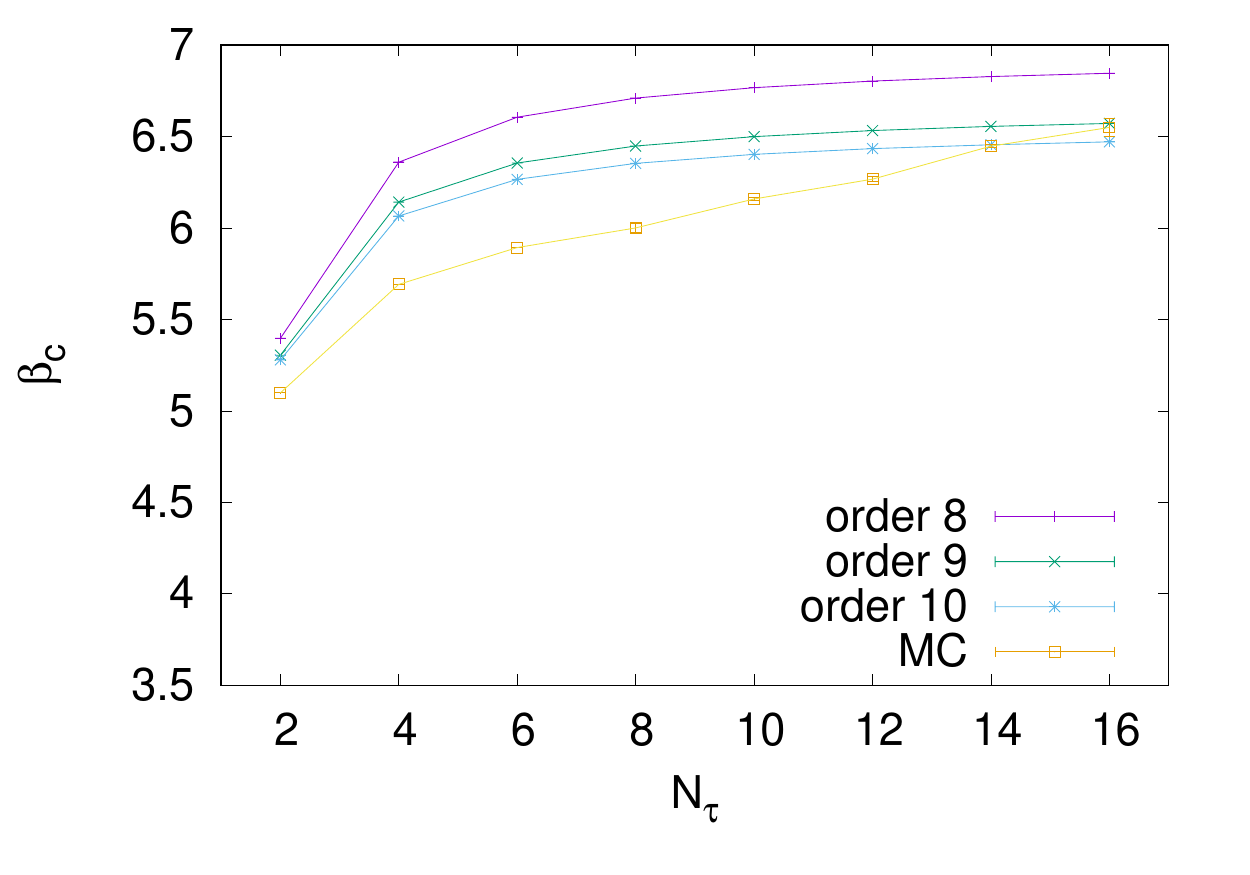}
\caption[]{ Left: Critical coupling of the effective lattice theory representing $SU(3)$ Yang-Mills.
Right: Critical Yang-Mills couplings predicted by the effective theory, compared to Monte Carlo results. From \cite{Kim:2019ykj}.}
\label{fig:efft_ym}
\end{figure}
\begin{figure}[t]
\centering
\includegraphics[width=0.45\textwidth]{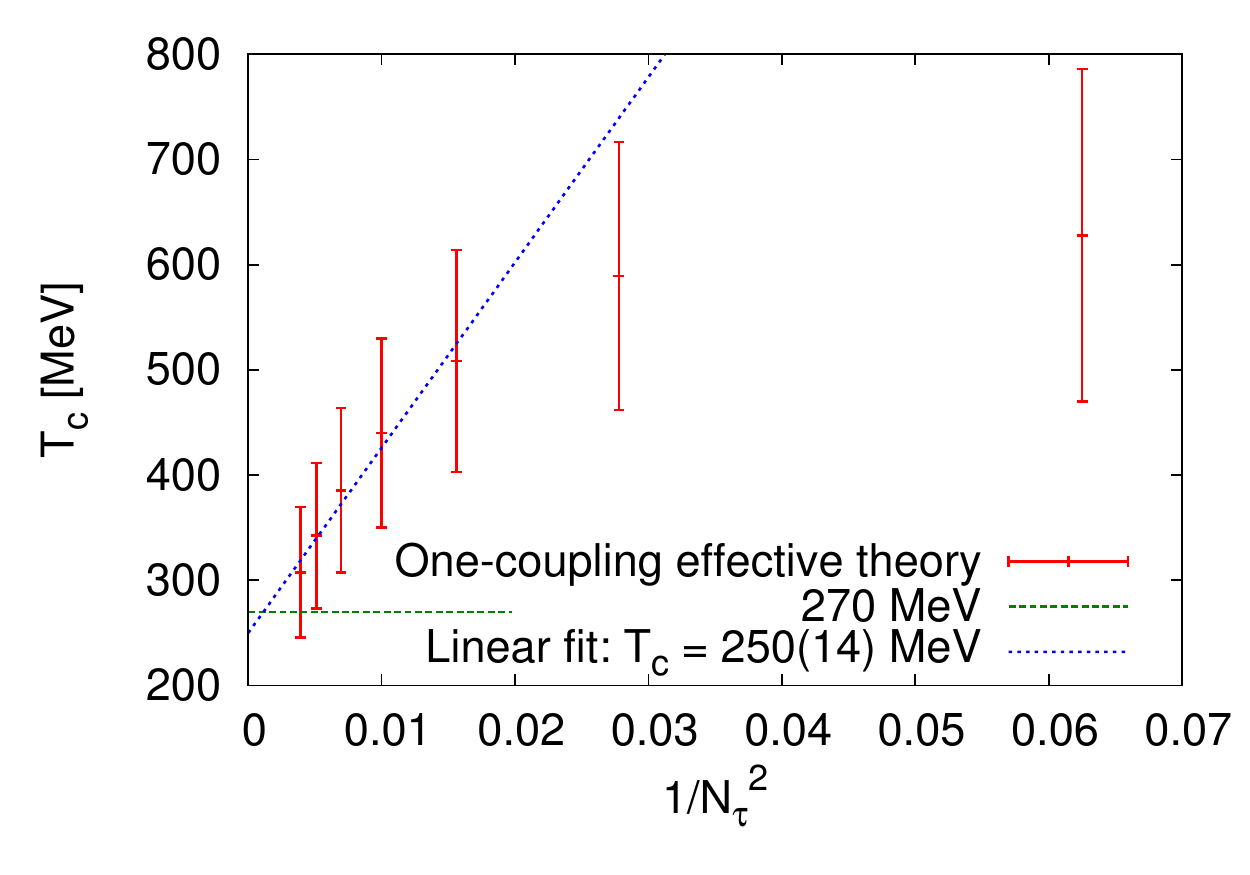}\hspace*{0.5cm}
\includegraphics[width=0.5\textwidth]{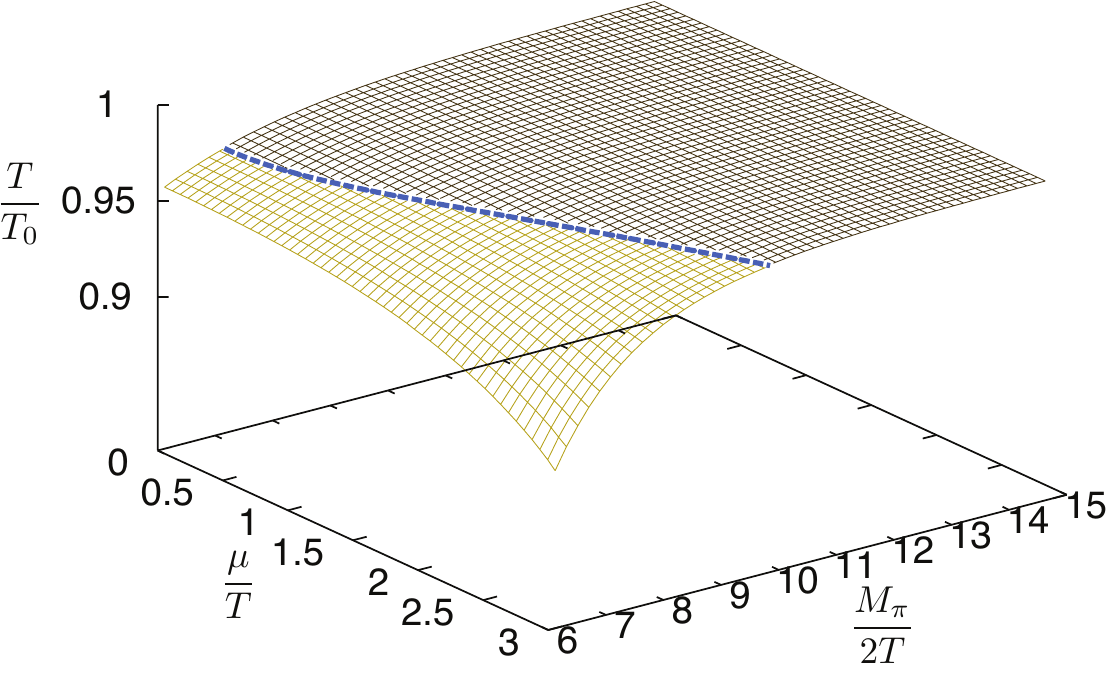}
\caption[]{ Left: 
Continuum extrapolation of critical temperature predicted by the effective theory.
Right: Phase diagram for QCD with heavy quarks. From  \cite{Fromm:2011qi}.}
\label{fig:efft_heavy}
\end{figure}

To appreciate the advantage of going through an effective theory, let us consider $SU(3)$ Yang-Mills theory, 
so that $h_1=\bar{h}_1=h_2=0$ and Eq.~(\ref{eq:zeff}) is reduced to the first line.
One can now compute the free energy and the susceptiblitiy of the Polyakov loop as a power series in $\lambda$, and improve the results
by Pad\'e approximants as in Sec.~\ref{sec:ym}. These show a singularity at a critical $\lambda_c$, indicating the phase transition. Note
however, that for $SU(3)$ we have a first-order phase transition, so that the singularity corresponds to the end of the metastability region
of the disordered phase, i.e., the true critical coupling is slightly overestimated. With a series through $O(\lambda^8)$ the estimates for $\lambda_c$
shown in Fig.~\ref{fig:efft_ym} (left) are obtained, which agree with a Monte Carlo determination within 2\%.  
This critical coupling can be converted into the critical Yang-Mills coupling $\beta_c(N_\tau)$ by inverting the first of Eq.~(\ref{eq:coupl}).
Comparison with the full 4d Monte Carlo result \cite{Fingberg:1992ju}
shows agreement to better than 10\% for the entire range of $N_\tau\in[2,16]$. This is a dramatic improvement compared to 
the direct calculations in 4d QCD in Sec~\ref{sec:ym}, which were only feasible for small $N_\tau$. This result also constitutes
a completely analytic calculation of the deconfinement transition of lattice Yang-Mills theory.
In fact,  the $\beta_c$-values for the larger $N_\tau$ are in the scaling region and a continuum extrapolation of the critical
temperature is possible, as shown in Fig.~\ref{fig:efft_heavy} (left). The resulting $T_c$ in the continuum also agrees within
10\% with the known value from full 4d simulations. This calculation demonstrates that, in some cases,
it is quite feasible to obtain continuum results from strong coupling methods.

\subsection{Deconfinement transition in QCD with heavy quarks}

Once quarks are included via the hopping expansion, a fully analytic evaluation  becomes much more complex due 
to the additional couplings, but the effective theory can still be simulated efficiently.
Moreover, while the partition function Eq.~(\ref{eq:zeff}) still has a sign problem, it is considerably milder
than that of 4d QCD, due to the fact that many degrees of freedom have been integrated out already.
In \cite{Fromm:2011qi} a flux representation of the effective theory including the static determinant was simulated,
and the resulting phase diagram for $N_f=2, N_\tau=6$ is shown in Fig.~\ref{fig:efft_heavy} (right).
Since the center symmetry of the Yang-Mills theory is explicitly broken by the quark determinant, the
first-order transition weakens with decreasing quark mass until it vanishes at some critical value $m_q^c$.
The transition also weakens with chemical potential, so that for $m_q>m_q^c$ one finds a first-order deconfinement
transition coming from the $T$-axis and terminating in a critical endpoint, whose location depends on the quark mass.
A continuum extrapolation would require larger $N_\tau$-values, which in turn demand higher orders
for the hopping expansion to converge. Note that recent 4d QCD simulations using a hopping expanded determinant \cite{Ejiri:2019csa},
as well as full QCD simulations \cite{Cuteri:2020yke} permit control over the systematics at $\mu=0$.
\begin{figure}[t]
\centering
\includegraphics[width=0.45\textwidth]{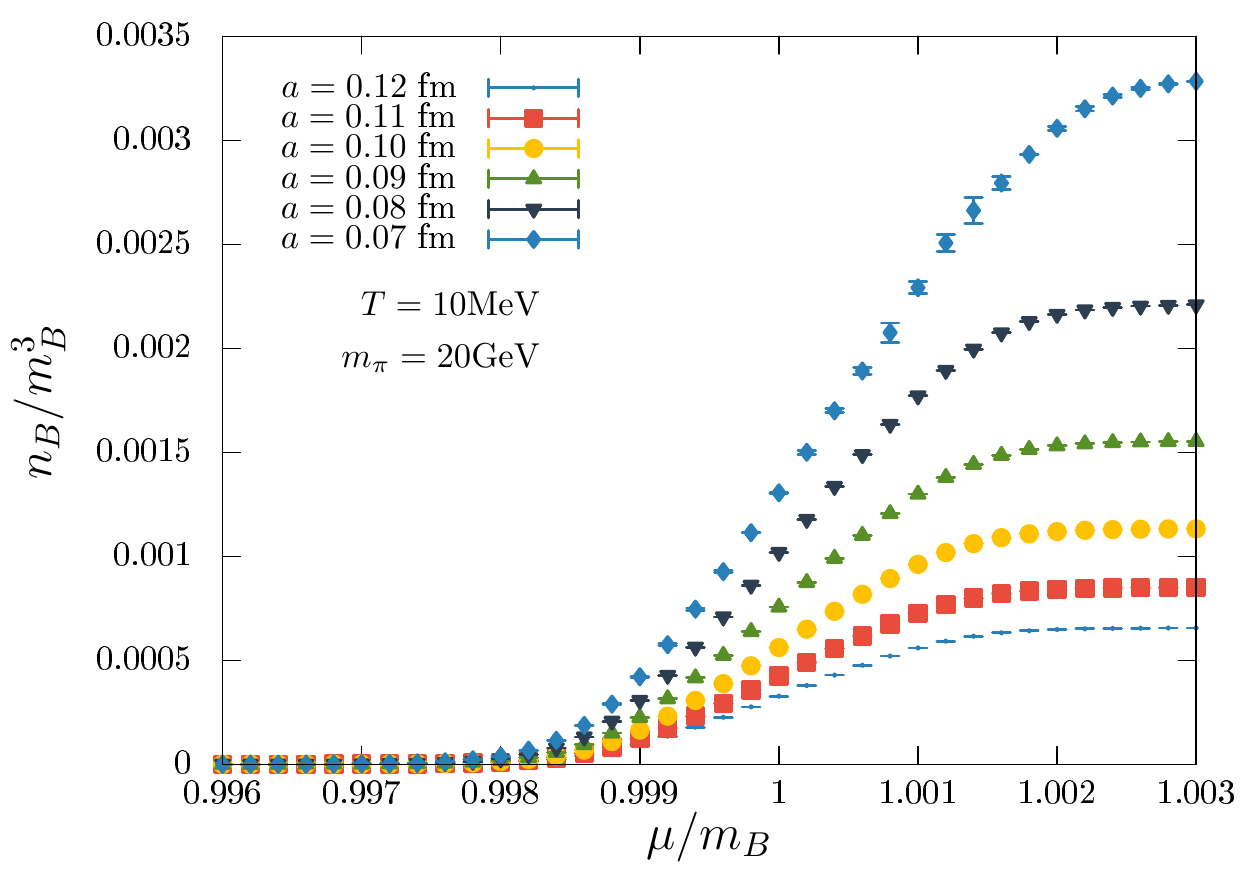}\hspace*{0.5cm}
\includegraphics[width=0.45\textwidth]{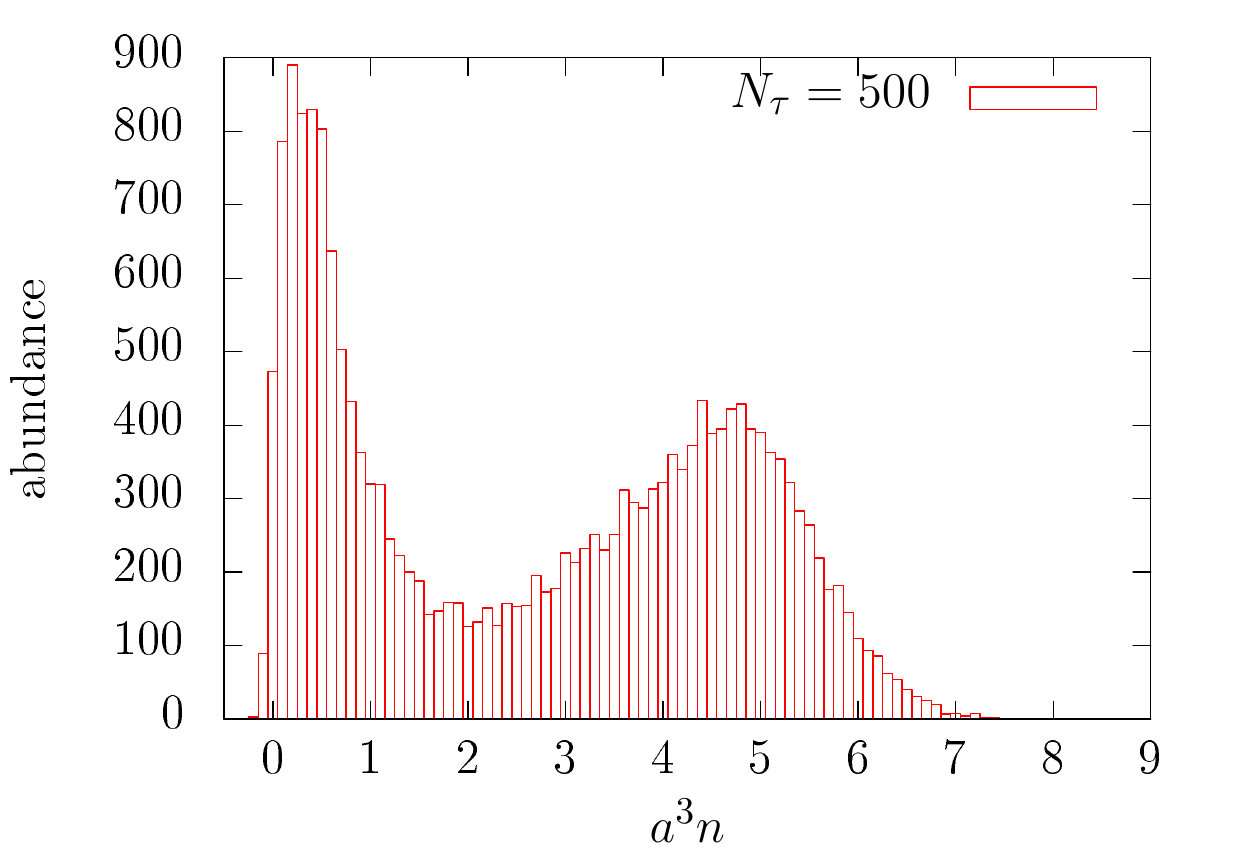}
\caption[]{Left: Onset of baryon number (crossover) for heavy quarks and different lattice spacings \cite{Glesaaen:2015vtp}.
Right: First-order onset transition for light quarks and low $T$ (large $N_\tau$) \cite{Langelage:2014vpa}.}
\label{fig:efft_dense}
\end{figure}

\subsection{Onset transition to baryon matter}

With the same methods the cold and dense regime, where the sign problem is most severe, 
can also be studied. In particular, the
onset transition to baryon matter has been seen explicitly 
to various orders in the combined expansions \cite{Fromm:2012eb,Langelage:2014vpa,Glesaaen:2015vtp}. 
Once more valuable physical insight can be gained by an analytic calculation to leading order within the effective theory.
In the strong coupling ($\beta=0$) limit with a static quark determinant only, the partition function 
factorises into one-site integrals which can be solved exactly. In the zero temperature limit,
mesonic contributions are exponentially suppressed by chemical potential and for  
$N_f=1$ we have 
\beq
Z(\beta=0) \stackrel{T\rightarrow 0}{\longrightarrow}z_0^V \quad \mbox{with}\quad 
z_0=1+4h_1^{3}+h_1^{6}, \quad h_1=(2\kappa e^{a\mu})^{N_\tau}=e^{(m-\mu)/T}\;.
\eeq
This corresponds to a free baryon gas with two species. With one quark flavour only, there are no nucleons and the
prefactor of the first term indicates a spin 3/2  quadruplet of $\Delta$'s, whereas the second term corresponds to 
a spin 0 six quark state or di-baryon.
The quark number density is now easily evaluated
\beq
n=
\frac{T}{V}\frac{\partial}{\partial \mu}\ln Z=\frac{1}{a^3}\frac{4N_ch_1^{N_c}+2N_ch_1^{2N_c}}{1+4h_1^{N_c}+h_1^{2N_c}}\;,
 \quad \lim_{T\rightarrow 0} a^3n=\left\{\begin{array}{cc} 0, & \mu<m\\
	2N_c, & \mu>m\end{array}\right.\;.
\eeq
Thus at $T=0$ we find a discontinous transition when the quark chemical potential equals the constituent mass $m$.
This reflects two expected physical phenomena: the ``silver blaze'' property of QCD, i.e.~the fact that the baryon number stays zero
for small $\mu$ even though the partition function explicitly depends on it \cite{Cohen:2003kd}. 
Once the baryon chemical potential 
$\mu_B=3 \mu$
is large enough to make a baryon ($m_B=3m$ in the static strong coupling limit), a discontinuous phase transition 
to a saturated baryon crystal takes place. 
The saturation density is $2N_c$ quarks per flavour and lattice
site and reflects the Pauli principle. This is clearly a discretisation effect that has to disappear
in the continuum limit.

In the case of two flavours the corresponding expression for the free gas of baryons reads
\begin{eqnarray}
z_0& =& (1 + 4 h_d^3 + h_d^6)+ (6 h_d^2 + 4 h_d^5) h_u+ (6 h_d + 10 h_d^4)h_u^2+ 
  (4 + 20 h_d^3 + 4 h_d^6)h_u^3 \nn \\
&&  + (10 h_d^2 + 6 h_d^5) h_u^4+ ( 4 h_d + 6 h_d^4) h_u^5 
  +(1 + 4 h_d^3 + h_d^6)h_u^6\;,
\label{eq:freegas}  
\end{eqnarray}
with now two $h_1$ couplings for the $u$- and $d$-quarks. In this case we easily identify in addition the 
spin 1/2 nucleons as well has many other baryonic multi-quark states with their correct spin degeneracy. A similar result is obtained for
mesons if we instead consider an isospin chemical potential in the low temperature limit \cite{Langelage:2014vpa}. 
Remarkably, the entire spin-flavour-structure
of the QCD bound states is obtained in this simple static strong coupling limit.

When corrections are added the step function is smoothed out. 
\fig\ref{fig:efft_dense} (left) shows the baryon density computed through orders $O(\kappa^8 u^5)$ and plotted in
continuum units. As the continuum is approached, the artefact of lattice saturation moves to infinity, as expected. 
\fig\ref{fig:efft_dense} (left) shows a crossover,
whereas for light quarks simulations show a first-order transition for large $N_\tau$, \fig\ref{fig:efft_dense} (right), and a crossover
for smaller $N_\tau$. This can again be understood with analytic insight from the hopping expansion.
The dimensionless quantity
\beq
\epsilon(\mu,T)= \frac{e-n_Bm_B}{n_Bm_B} =  -\frac{4}{3}\frac{1}{a^3n_B}\left(\frac{6h_1^3+3h_1^6}{z_0}\right)^2\,\kappa^2+\ldots\;,
\label{eq:bindpt}
\eeq
here evaluated at $\beta=0$ in the simpler case $N_f=1$  \cite{Langelage:2014vpa}, 
gives an interaction energy per baryon in units of the baryon mass when $T\rightarrow 0$. For $\mu<m$, it evaluates to 
zero in accordance with the silver blaze property. For $\mu>m$ it is nonzero and implies an attractive interaction,
consistent with the baryon gas `condensing` to a liquid (or crystal on the lattice).
Starting as $\sim\kappa^2$, it decreases with growing quark mass to zero in the static limit, 
as one would also expect from Yukawa potentials in nuclear physics.
Hence, the end point of the liquid gas transition, $T_c(m)$, shrinks to zero with increasing quark mass, which explains
the difference between the left and right of Fig.~\ref{fig:efft_dense}.   

\subsection{Large $N_c$ and quarkyonic matter}

\begin{figure}[t]
\centering
\includegraphics[width=0.45\textwidth]{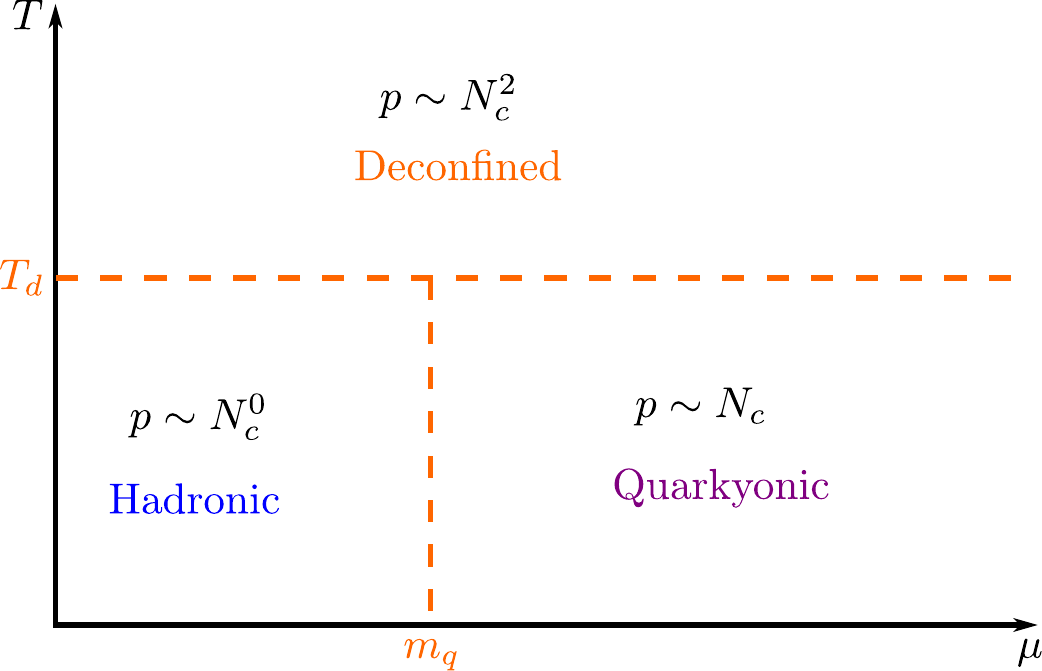} \hspace*{0.5cm}
\includegraphics[width=0.45\textwidth]{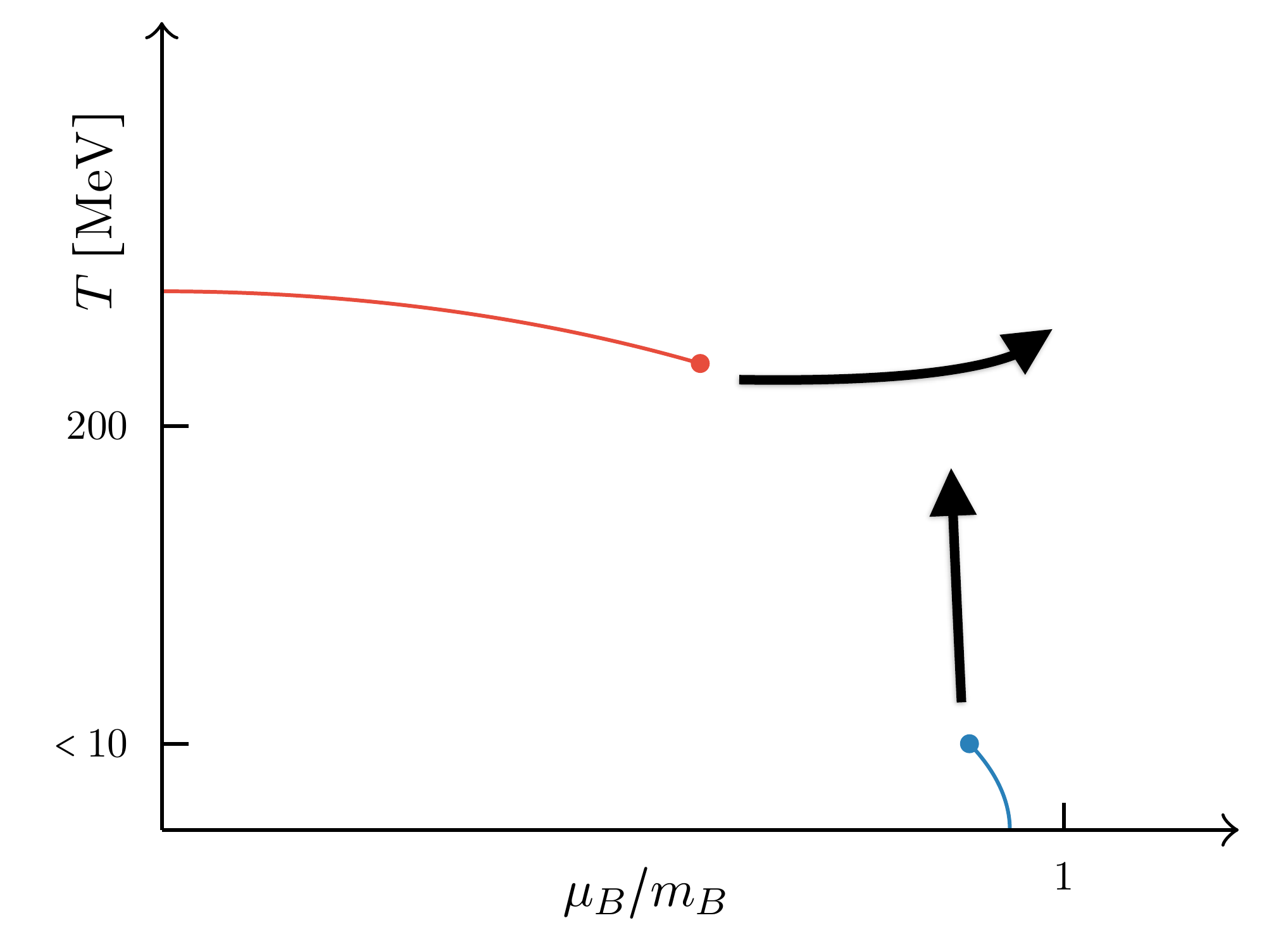}
\caption[]{Left: Phase diagram for QCD with $N_c\rightarrow \infty$ \cite{McLerran:2007qj}. 
Right: Evolution of the phase diagram for heavy quarks with growing $N_c$ \cite{Philipsen:2019qqm}.}
\label{fig:nc_pd}
\end{figure}
Interesting conjectures concerning the QCD phase structure were made some years ago,
based on large $N_c$ arguments \cite{McLerran:2007qj}. In particular, the phase diagram in the limit of large $N_c$, with fixed 
't Hooft coupling,
\beq
N_c\rightarrow \infty \quad \mbox{with}\quad  \lambda_H=g^2 N_c=\mathrm{const.},
\eeq
was argued to
look as in \fig\ref{fig:nc_pd} (left). With fermion contributions suppressed by large $N_c$, the deconfinement transition is a straight line
separating the plasma phase, where the pressure scales as $p\sim N_c^2$ (perturbation theory), 
from the hadron gas phase, where it scales as $p\sim N_c^0$ (ideal gas of hadrons).
In \cite{McLerran:2007qj} it is argued that at finite density there should then be a third phase with $p\sim N_c$, which was termed
quarkyonic since it shows aspects of both baryon and quark matter. In particular, at low temperatures the fermi sphere in momentum 
space is argued to be composed of a baryonic shell of thickness $\sim \Lambda_{QCD}$, and quark matter inside.

With the analytic expansion tools and an effective lattice theory at hand, one can 
now adress these issues by direct calculations starting from QCD.  
For large $N_c$, the baryon mass grows as $m_B\sim N_c$, so the size of the constituent quark mass should not matter in that limit.
This suggests to investigate the cold 
and dense region for large $N_c$ by direct calculation with the effective theory for heavy QCD. 
First, the effective theory of the previous sections was derived for a general number of colours~\cite{Christensen:2013xea} (see also
\cite{Borisenko:2020prf}).
Second, the onset transition to baryon matter has been analysed for general $N_c$ in \cite{Philipsen:2019qqm}, where
no large $N_c$ approximations are implied.  

\begin{figure}[t]
\centering
\includegraphics[width=0.45\textwidth]{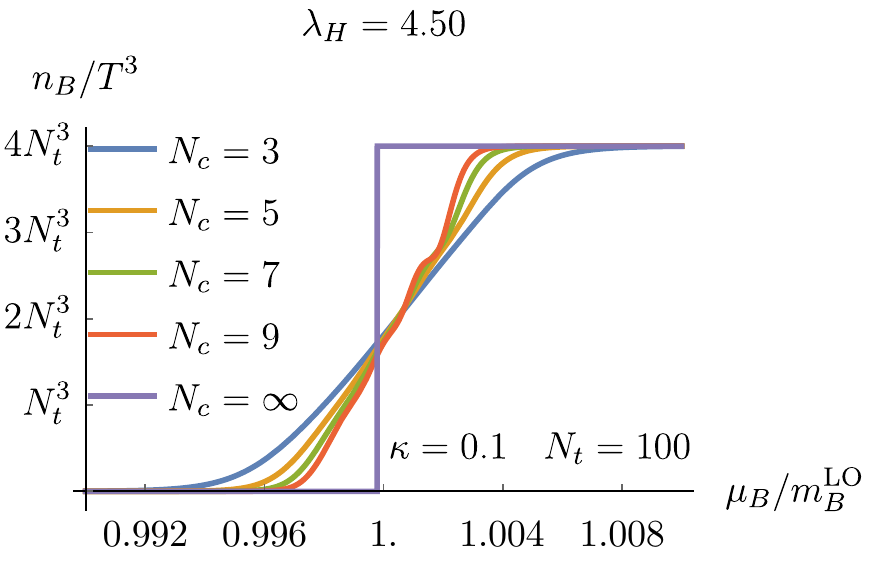} \hspace*{0.5cm}
\includegraphics[width=0.45\textwidth]{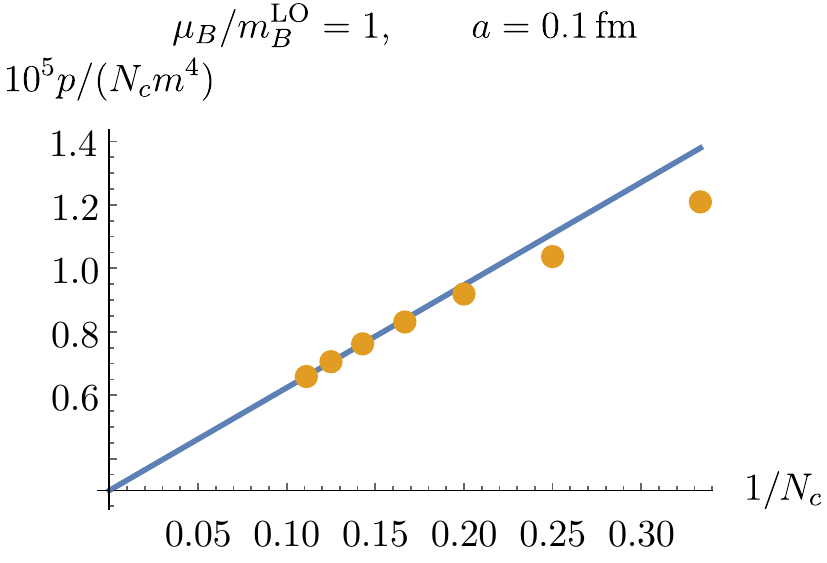}
\caption[]{Left: The onset transition always becomes first order
for large $N_c$. \\
Right: $p\sim N_c$ for large $N_c$ in baryon matter \cite{Philipsen:2019qqm}.}
\label{fig:nc}
\end{figure}
 \fig\ref{fig:nc} (left) shows the baryon onset transition to steepen with growing $N_c$, such that it always become first-order in the large $N_c$ limit
when the t'Hooft coupling and $N_\tau$ are held fixed. This is independent of the value of $N_\tau$, which parametrises temperature.
This strenghtening of the onset transition implies its critical endpoint $T_c(N_c)$ to move to ever larger temperatures with growing $N_c$,
as indicated in \fig\ref{fig:nc_pd} (right). Together with the straightening of the deconfinement line with $N_c$, which follows from
the behaviour of Feynman diagrams at large $N_c$, 
we thus understand how the conjectured phase diagram in \fig\ref{fig:nc_pd} (left) emerges smoothly
from the phase diagram with $N_c=3$ and heavy quarks. 
Furthermore, right after the onset transition $h_1\gsi 1$, 
through three orders in the hopping expansion at $\beta=0$, the coefficients of the pressure are proportional to $\sim N_c$
once $N_c$ is large, which is suggestive of a property to all 
orders. This $N_c$-scaling is stable under gauge corrections and, with a leading order correction in 
$N_c$, holds all the way down to $N_c=3-9$, \fig\ref{fig:nc} (right). 
Note also, that a lattice filling with baryon number smoothly changes from baryon matter (at the onset of condensation) 
to quark matter (at saturation) as a function of $\mu_B$, which is consistent with the momenum space picture of quarkyonic matter.

It thus seems that lattice QCD with heavy quarks shows the defining features
ascribed to quarkyonic matter. On the other hand, for the moderate densities right after the baryon onset, one
would expect no discernible difference from ordinary baryon matter. It remains an open question what happens at larger
densities, in particular 
for the case of light quarks, when there  
may or may not be an additional chiral transition after baryon onset.

\section{Effective lattice theory for chiral and light quarks}

The light quark regime has also been addressed already a long time ago by
effective theories at strong coupling. In particular, it was attempted to describe dense baryonic matter
in Hamiltonian approaches, with constructions based on the strong coupling limit and partial chiral symmetry on the 
lattice \cite{Bringoltz:2002qc,Bringoltz:2003jf,Bringoltz:2006pz}.
Here, we stick to the Euclidean approach, starting from the partition function, to apply systematic expansions.
However, being interested in the chiral phase transition, we now consder the lattice action with staggered fermions,
\begin{eqnarray}
\label{Z_standard}
\hspace*{-0.75cm}
Z &=& \prod\limits_{x}\int d\chi_{x}d\bar{\chi}_{x}e^{2am_{q}\bar{\chi}_{x}\chi_{x}}\prod_{\ell}\int dU_{\ell}\;e^{-S_\mathrm{YM}[U]}\cdot e^{\tr \left[U_{\ell}M_{\ell}^{\dag} + U^{\dag}_{\ell}M_{\ell}\right]}, \\
\left(M^{\dag}\right)_{i}^{j} &=&  \eta_{\mu}(x)e^{a\mu_{B}\delta_{\mu,0}}\bar{\chi}_{x}^{i}\chi_{x+\mu,j},\qquad  M_{k}^{l} = -\eta_{\mu}(x)e^{-a\mu_{B}\delta_{\mu,0}}\bar{\chi}_{x+\mu}^{k}\chi_{x,l},\nn
\end{eqnarray}
where $(\ell, x, p)$ label lattice links, sites and plaquettes and $\eta_\mu(x)$ are the staggered phases. 
In the continuum limit, this represents $N_f=4$ QCD 
if no rooting is applied.
Expanding in $\beta$, one obtains
\begin{eqnarray}
Z&=&\prod\limits_{x}\int d\chi_{x}d\bar{\chi}_{x}\,\,e^{2am_{q}\bar{\chi}_{x}\chi_{x}}\\
&&
\sum_{\{n_{p},\bar{n}_{p}\} }\!\prod_{\ell,p}\frac{\left(\frac{\beta}{2N_c}\right)^{n_{p}+\bar{n}_{p}}}{n_{p}!\bar{n}_{p}!}\!\!\int dU_{\ell}\tr[U_{p}]^{n_{p}}\tr[U^{\dag}_{p}]^{\bar{n}_{p}}e^{\tr \left[U_{\ell}M_{\ell}^{\dag} + U^{\dag}_{\ell}M_{\ell}\right]}\;, \nonumber
\end{eqnarray}
with plaquette (anti-plaquette) occupation numbers $\{n_{p},\bar{n}_{p}\}$. 
In the strong coupling limit, $\beta=0$, the gauge action vanishes and the link integration reduces to one-link integrals \cite{Eriksson:1980rq}.
Subsequent integration over the Grassmann variables then yields the partition function as a sum over graphs in terms
of mesons and baryons \cite{Rossi:1984cv}. Leading $O(\beta)$-corrections have been derived in \cite{deForcrand:2014tha}.
Recently, after an intricate reorganisation of the involved degrees of freedom, an all-order dual formulation 
in terms of monomers, dimers, world lines and world sheets has been given
\cite{Gagliardi:2017uag,Gagliardi:2019cpa}. Its contributions through $O(\beta^2)$ can be simplified to  
\bea
Z(m_q,\mu)&=&\sum_{\{k,n,l,n_p\}}\underbrace{\prod_{b=(x,\nu)}\frac{(N_c-k_b)!}{N_c!(k_b-|f_b|)!}}_{\tiny\mbox{meson hoppings}}
\underbrace{\prod_x\frac{N_c!}{n_x!}(2am_q)^{n_x}}_{\tiny\mbox{chiral condensate}}\nn\\
&&\hspace*{1.5cm}\times \underbrace{\prod_{l_3}w(l_3,\mu)\prod_{l_f} \tilde{w}(l_f,\mu)}_{\tiny\mbox{baryon hoppings}}
\underbrace{\prod_p\frac{(\frac{\beta}{2N_c})^{n_p+\bar{n}_p}}{n_p!\bar{n}_p!}}_{\tiny\mbox{gluon propagation}}\;.
\eea
The integer valued occupation variables  satisfy constraints at sites and bonds, which result from the Grassman and gauge integrations.
Note that this formulation in general also contains negative weights, but the remaining sign problem is mostly mild enough to be handled
by reweighting techniques. A particular advantage of this formulation is the feasibility to simulate the chiral limit directly, as well as
any finite quark mass. On the other hand, gauge corrections are more difficult to include. Moreover, in the present formulation
the expansion of
the gauge sector in $\beta$ is slower to converge than the previously discussed
character expansion, to which it might be extended in the future. 
Nevertheless, this formulation offers yet another road to studying QCD at finite density in a complementary
parameter regime compared to the previous sections.
\begin{figure}[t]
\centering
\vspace*{-0.3cm}
\includegraphics[width=0.5\textwidth]{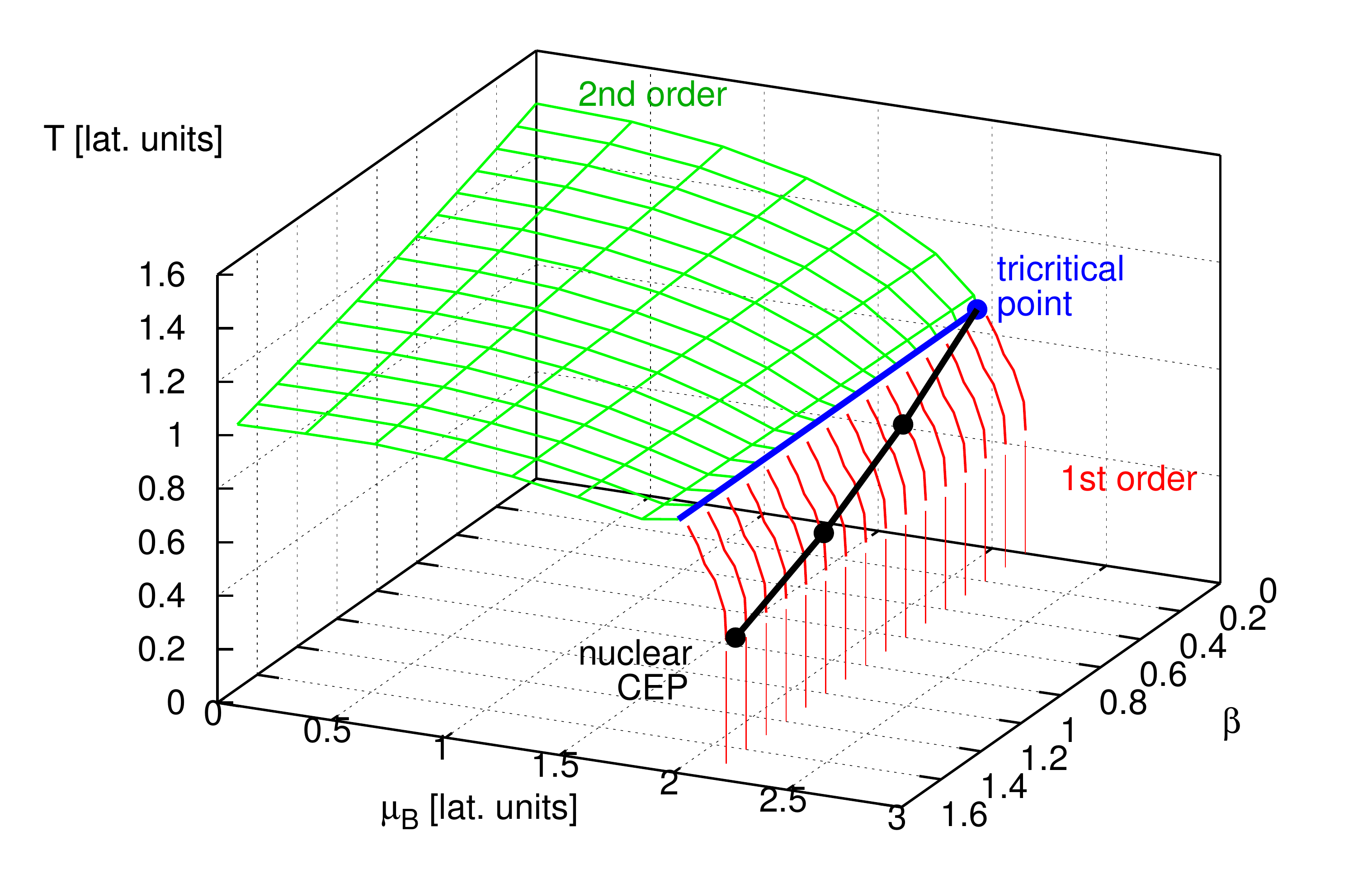}\hspace*{0.5cm}
\includegraphics[width=0.4\textwidth]{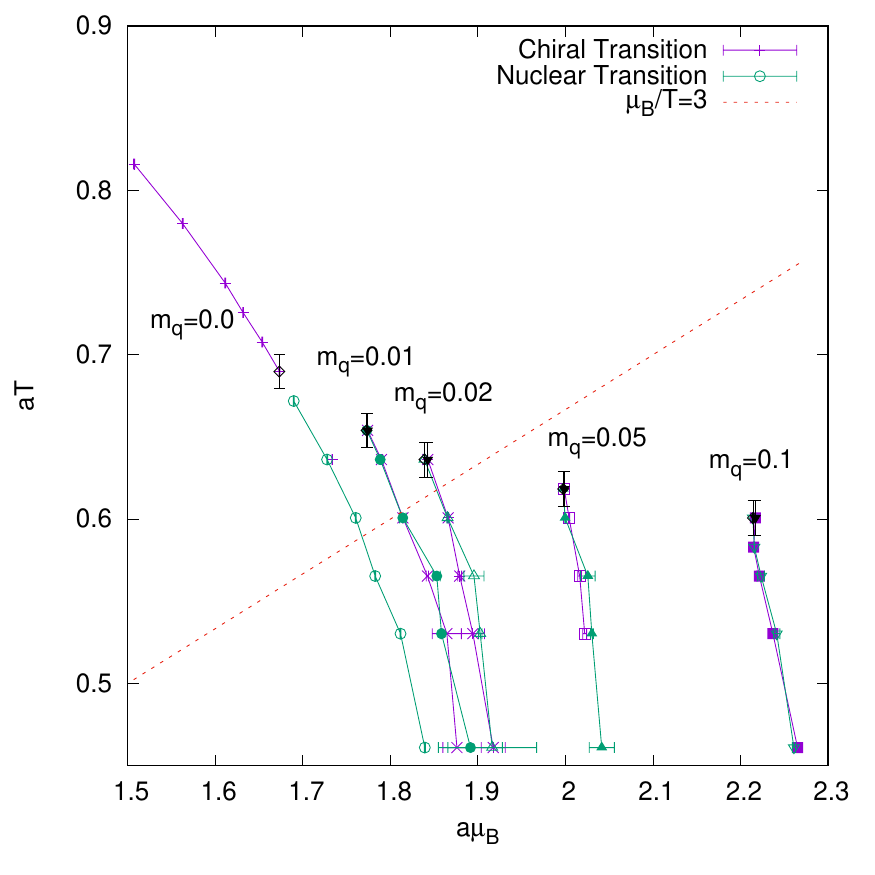}
\caption[]{Left: Chiral phase transition and onset transition to baryon matter for $m=0$ \cite{deForcrand:2014tha}.\\ 
Right: Mass dependence of the chiral and baryon
onset transition at $\beta=0$ \cite{Kim:2016izx}.}
\label{fig:efft_chiral}
\end{figure}

Early mean field \cite{Kawamoto:1981hw} and Monte Carlo \cite{Karsch:1988zx} studies based on a polymer representation
have been restricted to the strong coupling limit, $\beta=0$. Modern simulations are done using a worm 
algorithm \cite{deForcrand:2009dh}, which can be extended to include gauge corrections.
\fig\ref{fig:efft_chiral} (left) shows the phase diagram in the chiral
limit, $m=0$, both for $\beta=0$ and with leading gauge corrections $O(\beta)$ included \cite{deForcrand:2014tha}. In the chiral limit there always 
is a non-analytic chiral phase transition, with 
a tricritical point where the first-order transition at finite density meets the second-order line. 
In the strong coupling limit, this tricritical point  coincides with the end point of the nuclear liquid gas transition. 

When gauge corrections are switched on,
these start splitting up, but surprisingly the first-order lines of the chiral and nuclear transitions are still indistinguishably close.
\fig\ref{fig:efft_chiral} (right) shows the strong coupling limit, but now with finite quark mass switched on. The second-order 
transition line changes to crossover, as expected. A very intersesting feature is the decrease with mass of the endpoint $T_c(m)$ of the 
nuclear liquid gas transition. Obtained from a different discretisation and in a very different parameter range,
this behaviour agrees with the finding for heavy quarks discussed in the last sections. Note also that the endpoint of the chiral transition  
quickly moves to $\mu_B\gsi 3T$, which is 
consistent with all results reported for QCD at the physical point \cite{Ratti:2019tvj,Philipsen:2019rjq}.

\section{Conclusions}

In view of the  sign problem of QCD at finite baryon density, which continues  to evade purely algorithmic solutions, 
systematic studies of strong coupling methods, either directly or through development of effective theories, have provided a number
of genuinely new results for QCD thermodynamics:  

\begin{itemize}
\item The behaviour of screening masses for temperatures up to the crossover
\item A derivation of the hadron resonance gas in the strong coupling regime
\item The deconfinement transition for heavy quarks at all baryon chemical potentials
\item The chiral phase transition in the massless limit in the region of strong, but finite, couplings
\item The onset transition to baryon matter and its equation of state, both for heavy quarks on fine lattices
and for light quarks on coarse lattices
\item A smooth connection to thermodynamics at large $N_c$  
\end{itemize}  

These results were so far obtained in parameter regions away from the physical point in the continuum. 
Nevertheless, all the physics expected up to nuclear densities has been seen qualitatively on the lattice. It is now 
a matter of pushing these approaches to higher orders, and to evaluate them with refined algorithms.  
Given existing automatisation techniques to achieve high-order calculations for spin models and their sign problems,
known in the condesed matter literature,
the author believes this path is worth to be pursued further. 

\begin{acknowledgements}
This article is dedicated to the memory of Pushan Majumdar. Pushan was my first postdoc when I built my first group as a starting 
professor at the Westf{\"a}lische Wilhelms-Universit{\"a}t M\"unster, Germany. 
Even though we never published a paper together, 
Pushan was an indispensable assistant in bouncing research ideas back and forth, and designing exercises and exams for my 
classes.
We were in constant scientific exchange,
be it about separate projects that each one "just needed to finish", or about the projects of a rapidly growing number of graduate students in the
group. In particular, Pushan co-supervised the Master thesis of Bastian Brandt on effective string theories in QCD, and he
vividly followed the beginning of the strong coupling journey 
described in this article. His genuine interest in physics, no matter how close or far from his own current research projects, 
greatly contributed to the scientific evolution of the group.
With his open personality and subtle humour he also shared in social activities, and will be remembered.

I thank W.~Unger for useful comments on the manuscript.
Part of the work reported here was supported by the Deutsche Forschungsgemeinschaft (DFG, German Research Foundation)
 -- project number 315477589 -- TRR 211.

\end{acknowledgements}

%
%

\newpage
\bibliographystyle{spphys}       
\bibliography{references}   

%
%

\end{document}